\documentclass[
 aps,
 prl,
 superscriptaddress,
 reprint,
 nofootinbib 
 ]{revtex4-1}
\usepackage[utf8]{inputenc} 
\usepackage{hyperref} 
\usepackage{amsmath} 
\usepackage{amsfonts}
\usepackage{amsthm}
\usepackage{amssymb}
\usepackage{graphicx} 
\usepackage{lmodern} 
\usepackage{braket} 
\usepackage{subcaption} 
\usepackage{color} 
\usepackage[normalem]{ulem} 

\graphicspath{{./}{./figs/}}

\newcommand{\hsq}{\Phi^\dagger \Phi}
\newcommand{\te}{\textemdash}

\newcommand{\tLO}{\text{LO}}
\newcommand{\tNLO}{\text{NLO}}
\newcommand{\tNNLO}{\text{NNLO}}
\newcommand{\tmin}{\text{min}}
\newcommand{\define}{\equiv}
\newcommand{\yc}{y_\text{c}}
\newcommand{\Tc}{T_\text{c}}
\newcommand{\sectioninline}[1]{\textit{#1}.---} 

\definecolor{OGcolour}{rgb}{0,0.5,0}

\newcommand{\Nottingham}{\affiliation{
		School of Physics and Astronomy,
		University Park,
		University of Nottingham,
		Nottingham NG7 2RD,
		United Kingdom
}}

\newcommand{\Uppsala}{\affiliation{
		Department of Physics and Astronomy, Uppsala University,
		Box 516, SE-751 20 Uppsala,
		Sweden
}}

\newcommand{\Hamburg}{\affiliation{
		II. Institute of Theoretical Physics, Universität Hamburg, D-22761, Hamburg, Germany
}}

\newcommand{\DESY}{\affiliation{
		Deutsches Elektronen-Synchrotron DESY, Notkestr. 85, 22607 Hamburg, Germany
}}

\begin{document}
	
	\title{Radiative first-order phase transitions to next-to-next-to-leading-order}
	\date{May 16, 2022}
	
	\author{Andreas Ekstedt}
	\email{andreas.ekstedt@desy.de}
	\Uppsala \Hamburg \DESY
	
	\author{Oliver Gould}
	\email{oliver.gould@nottingham.ac.uk}
	\Nottingham
	
	\author{Johan L\"{o}fgren}
	\email{johan.lofgren@physics.uu.se}
	\Uppsala

\begin{abstract}
We develop new perturbative tools to accurately study radiatively-induced first-order phase transitions. Previous perturbative methods have suffered internal inconsistencies and been unsuccessful in reproducing lattice data, which is often attributed to infrared divergences of massless modes (the Linde problem). We employ a consistent power counting scheme to perform calculations, and compare our results against lattice data. We conclude that the consistent expansion removes many previous issues, and indicates that the infamous Linde problem is not as big a factor in these calculations as previously thought.
\end{abstract}

\maketitle


\sectioninline{Introduction}%
 \fontdimen2\font=0.6ex
A cosmological first-order phase transition would have far-reaching consequences for the early universe. Not only could the transition trigger electroweak baryogenesis  \cite{Kuzmin:1985mm}, and thereby explain the matter-antimatter asymmetry, but the violent transition would echo across the universe; leaving a stochastic background of gravitational waves in its wake. This makes searching for the telltale spectral shape \cite{Hindmarsh:2017gnf} a key scientific objective of current and planned gravitational wave observatories \cite{NANOGrav:2020bcs, LISA:2017pwj, DECIGO:2011zz, Taiji:2018tsw, AEDGE:2019nxb}. Besides giving clear evidence for a first-order transition, a detected signal would constrain models of dark matter \cite{Schwaller:2015tja, Croon:2018erz}, baryogenesis \cite{Vaskonen:2016yiu}, inflation \cite{Addazi:2018nzm} and grand unification \cite{Croon:2018kqn}.

Yet current tools have a hard time making reliable predictions, particularly so for phase transitions which are \emph{radiatively induced},%
\footnote{We here use the term ``radiative'' to denote radiative corrections from energy scales much lower than the temperature.}
where perturbative calculations often cannot distinguish between first-order, second-order, and crossover transitions. Alternatively, lattice Monte-Carlo simulations can provide quantitatively reliable predictions, up to (small) statistical uncertainties. But they are slow, typically requiring thousands of CPU hours to accurately determine the thermodynamics of a single parameter point.

As a consequence, even with its problems, perturbation theory is the only tool capable of scanning the high dimensional parameter spaces of theories beyond the Standard Model. So it is imperative to put perturbative calculations on a solid footing.

In this paper we vie to do just that.
In particular, we construct a new perturbative expansion for models with radiatively-induced phase transitions. Furthermore, to ensure accuracy we compare the results with lattice data.  In the process we draw attention to prevailing problems with standard perturbative approaches in this context (for an overview see Ref.~\cite{Croon:2020cgk}).

\sectioninline{Theoretical challenges}%
Consistent calculations should be renormalization-scale invariant~\cite{Gould:2021oba}, gauge invariant \cite{Buchmuller:1994vy, Hirvonen:2021zej, Lofgren:2021ogg}, and free from infrared (IR) divergences~\cite{Linde:1980ts, Laine:1994bf, Laine:1994zq}.

In the present context, most significant renormalization-scale dependence originates from an hierarchy between the temperature and particle masses. Such an hierarchy is typically present in weakly coupled theories because large temperatures are needed for thermal loop-corrections to overpower the tree-level potential. And so, because different loop orders are mixed, the expansion is reorganized. For example, two-loop calculations are required to achieve any parametric cancellation of the scale dependence \cite{Gould:2021oba}.
Additionally, large logarithms arise due to the hierarchy of scales. Both these issues can be remedied by working with a dimensionally-reduced effective field theory (EFT) \cite{Farakos:1994kx, Kajantie:1995dw, Braaten:1995cm, Braaten:1995jr}.
 
This hierarchy of scales also leads to a gauge dependence problem, which can be resolved in the same manner when there is a tree-level barrier in the dimensionally reduced EFT \cite{Croon:2020cgk, Gould:2021oba}.
The gauge dependence problem is more sinister for radiatively induced transitions, where standard methods yield IR divergences \cite{Laine:1994bf, Laine:1994zq}.

We study first-order phase transitions in the three-dimensional (3d) $\mathrm{SU}(2)$ Higgs EFT. This model has been extensively studied using lattice simulations \cite{Kajantie:1995kf, Gurtler:1997hr, Rummukainen:1998as, Moore:2000jw, Gould:2022ran}, due to its relevance for the electroweak phase transition. The availability of lattice data allows us to determine the quantitative reliability of perturbation theory as a function of the expansion parameter. We also show how the perturbative expansion converges by pushing calculations to next-to-next-to-leading order (NNLO).

The 3d EFT that we study describes phase transitions of many extensions of the electroweak sector.
It has been used to describe phase transitions in the Minimally Supersymmetric Standard Model \cite{Laine:1996ms, Cline:1996cr, Farrar:1996cp, Cline:1997bm}, the two Higgs doublet model \cite{Losada:1996ju, Andersen:1998br, Andersen:2017ika, Gorda:2018hvi}, the singlet scalar extension of the Standard Model \cite{Brauner:2016fla, Gould:2019qek}, and the real triplet scalar extension of the Standard Model \cite{Niemi:2018asa}. The model also serves as a useful prototype for models with a radiatively-induced barrier.

\sectioninline{A modified expansion}%
This paper focuses on the 3d EFT, which contains all the phase-dependent physics of the transition. The exact form of the four-dimensional (4d) theory is not of interest. The action of the EFT,
barring gauge-fixing terms, is
\begin{align}
    S_3=\int d^3x\left[\frac{1}{4} F_{ij}^a F_{ij}^a + D_i \Phi^\dagger D_i \Phi +V(\Phi)\right],
\end{align}
where $i,j$ run over the spatial indices and $a$ runs over the adjoint indices of $\mathrm{SU}(2)$. The notation is standard and follows Ref.~\cite{Farakos:1994kx}. The gauge coupling squared is denoted $g_3^2$.
The tree-level potential is
\begin{align}
    V(\Phi)=m^2_3 \hsq +\lambda_3 (\hsq)^2.
\end{align}
All three parameters, $g_3^2$, $m^2_3$ and $\lambda_3$ are effective\te temperature dependent\te parameters. The couplings are related to their 4d counterparts, and the temperature $T$, as $g_3^2\approx g^2T$ and $\lambda_3\approx \lambda T$. The parameter $m^2_3$ is positive at high temperatures and negative at low temperatures\te schematically $m^2_3 \approx - \alpha + \gamma T^2 $ for positive constants $\alpha$, $\gamma $.

The phase structure of the model can be studied perturbatively via the effective potential. As such we make the usual expansion, $\Phi \to (0,\tfrac{1}{\sqrt{2}}\phi) + \Phi$, where $\phi$ denotes the vacuum expectation value (vev). In general $\phi$ can either be zero (symmetric-phase) or non-zero (broken phase). These phases have different free energy densities; which, when overlapping, identifies a first-order phase transition.

Looking at the tree-level potential in terms of $\phi$, it would appear that no first-order transition, with a barrier between coexisting phases, is possible. Yet a barrier can be generated by vector-bosons loops~\cite{Arnold:1992rz}. Indeed, incorporating the one-loop vector-boson diagram
gives an effective potential
\begin{align}\label{eq:BarrierPotential}
    V_\text{eff}(\phi)=\frac{1}{2}m_3^2\phi^2-\frac{g_3^3}{16\pi}|\phi|^3+\frac{\lambda_3}{4}\phi^4+\ldots
\end{align}
Eyeballing equation \eqref{eq:BarrierPotential}, the symmetric minimum ($\phi=0$) has lowest energy for large $m_3^2$; while the broken minimum ($\phi\neq 0$) has lowest energy for small $m_3^2$. These two minima overlap for some intermediate $m_3^2$ value. This occurs when terms in the potential are of similar size~\cite{Arnold:1992rz}:
\begin{align}
    \phi&\sim \frac{g_3^3}{\lambda_3},&
    m_3^2&\sim\frac{g_3^6}{\lambda_3}.\label{eq:scaling1}
\end{align}
Therefore, the vector-boson mass $m_A^2=\tfrac{1}{4}g_3^2\phi^2$ and the scalar-boson mass are related by $m_3^2/m_A^2\sim \lambda_3/g_3^2$.

Then, if this ratio,
\begin{align} \label{eq:x}
    x\define\frac{\lambda_3}{g_3^2},
\end{align}
is parametrically small we can formally integrate out the vector bosons, which is what generates the barrier in equation \ref{eq:BarrierPotential}. Because of this, the perturbative expansion is performed in powers of $x$.

Note that in counting powers of $x$, one can directly identify the (potentially infinite) classes of Feynman diagrams which contribute at a given order in $x$, and make appropriate resummations, and subtractions, to avoid double counting \cite{Arnold:1992rz, Ekstedt:2020abj}. Alternatively, as we do here, one can use EFT-techniques to systematically integrate out vector bosons. When expanded strictly in $x$, both approaches give the same results.

Within this EFT, perturbation theory should work well for small $x$. However, for sufficiently small $x$ the vector mass scales as: $m_A \sim T$. This implies that the high-temperature expansion is invalid, and hence the original 3d EFT no longer faithfully describes the infrared physics of the 4d theory. While for $x\sim 1$ vector boson masses are of the same order as scalar ones, implying that we can not consistently integrate them out. In this case, the dynamics of the transition is nonperturbative, and lattice simulations show either a second-order or crossover transition \cite{Gurtler:1997hr, Rummukainen:1998as}.

Focusing on first-order transitions, we consider $x\ll 1$ exclusively, with the caveat that our model can only be matched to a 4d theory if $x\approx \frac{\lambda}{g^2}\sim g/(4\pi)$.

In addition to $x$, we use $y \define \frac{m_3^2}{g_3^4}$ \cite{Kajantie:1995kf}. With these dimensionless variables, and scaling $\phi\to g_3 \phi$, the leading-order potential is
\begin{align}\label{eq:LeadingOrderPotential}
   \frac{V_\tLO(\phi)}{g_3^6}=\frac{1}{2}y\phi^2-\frac{1}{16\pi}|\phi|^3+\frac{x}{4}\phi^4.
\end{align}
In the broken phase, this is of order $\sim x^{-3}$.

In principle $|y|^{-1/2}$ could also act as an expansion parameter, as it appears in loop corrections through Feynman integrals. However, close to the critical temperature $y$ scales as $y\sim x^{-1}$, so it is enough to count powers of $x$. Hereafter we set $g_3^2=1$; if necessary, factors of $g_3^2$ can be reinserted by dimensional analysis.

In the EFT approach, one first integrates out the heavy vector fields, to create an EFT for the light scalar fields \cite{Gould:2021ccf, Lofgren:2021ogg, Hirvonen:2021zej}. At LO this gives Eq.~\eqref{eq:LeadingOrderPotential}. Subleading corrections to the EFT action from the heavy vector fields appear at NLO in the $x$ expansion. These corrections come with integer powers of $x$. After vector bosons are integrated out, the resulting EFT only contains scalars. These scalars give loop corrections which are suppressed by powers of $\lambda_3/m_3 \sim x^{3/2}$.

Thus, in the full coupled gauge-Higgs theory, the perturbative expansion about equation \eqref{eq:LeadingOrderPotential} is a dual expansion in powers of $x$ and $x^{3/2}$ (up to logarithms), and should be reliable when $x\ll 1$.

That said, the potential to NNLO reads
\begin{align}
   V_\text{eff}=V_\tLO+ x V_\tNLO + x^{3/2} V_\tNNLO +\ldots \label{eq:Veffx}
\end{align}
Here factors of $x$ only signify the suppression of higher-order terms. Importantly, the expansion is organized in powers of $x$\te not by loops.

This $x$-expansion describes radiatively-induced first-order phase transitions in an IR-finite and gauge invariant manner. The IR-finiteness of the $x$-expansion should be contrasted with the IR divergences which occur at two-loop order and above in a direct loop/$\hbar$-expansion \cite{Laine:1994bf, Laine:1994zq}. The importance of the $x$-expansion has been long recognised \cite{Kajantie:1997tt, Kajantie:1997hn, Moore:2000jw}, yet in this paper for the first time we push the calculations to NNLO \te the lowest order at which resummations are necessary within 3d. As a consequence we are able to elucidate generic properties of the expansion.

Achieving order-by-order gauge invariance for observables requires the $x$-expansion to be performed strictly, so that all quantities are expanded in $x$, including intermediate quantities such as the Higgs vev \cite{Nielsen:1975fs, Fukuda:1975di}.


As mentioned, the next-to-leading order (NLO) potential comes from integrating out vector-bosons at two loops. While contributions from the scalar fields appear first at NNLO. These should be computed within the effective description for the light scalar fields. Thus, at LO the squared masses of the Higgs and Goldstone fields are, respectively,
\begin{align} \label{eq:resummed_masses}
    m_H^2(\phi)&= \partial^2_\phi V_\tLO,& m_G^2(\phi)&= \phi^{-1}\partial_\phi V_\tLO.
\end{align}
Utilising the LO potential rather than the tree-level potential here resums the Higgs and Goldstone self-energies to LO in $x$.


Now, all quantities, including the minima, should be expanded in powers of $x$. For example, the minimization condition is
\begin{align}\label{eq:Minimization}
    \partial_\phi V_\text{eff}(\phi)\left.\right|_{\phi=\phi_\tmin}=0,
\end{align}
where
\begin{align}\label{eq:HbarExpansion}
    \phi_\tmin=\phi_\tLO+x \phi_\tNLO+\mathellipsis,
\end{align}
and $\phi_\tLO$ solves $\partial_\phi V_\tLO(\phi)\left.\right|_{\phi=\phi_\tLO}=0$. Higher-order terms of $\phi_\tmin$ are found by using equation~\eqref{eq:HbarExpansion} in a Taylor expansion of equation~\eqref{eq:Minimization}. 

The effective potential evaluated at a minimum represents the energy of that phase. And the difference in free energy density between phases can be expressed as
\begin{align}
   \Delta V\define \left[V_\text{eff}(\phi_\tmin)-V_\text{eff}(0)\right]
\end{align}

We say that a phase-transition occurs for some critical mass, or value of $y$, defined by $\Delta V(y=\yc)=0$. From this one can determine the critical temperature $\Tc$ by solving $y(\Tc)=\yc$, and using the known temperature dependence of $y$ for a given 4d model.
This critical mass should also be found order-by-order in $x$, to wit $\yc=y_\tLO+x y_\tNLO+\ldots$. To leading order the critical mass is the solution of $\Delta V_\tLO=\left[V_\tLO(\phi_\tLO)-V_\tLO(0)\right]_{y=y_\tLO}=0$. And the next-to-leading order critical mass is
\begin{align}
    y_\tNLO=-\left.\frac{\Delta V_\tNLO}{ \partial_y \Delta V_\tLO}\right\vert_{\phi=\phi_\tLO,y=y_\tLO}.
\end{align}
Consider now an observable, $F(\phi,y)=F_\tLO+ x F_\tNLO+\ldots$, evaluated at the critical mass. The expansion is of the form
\begin{align}\label{eq:ObservableExp}
   & F(\phi_\tmin,\yc)=F_\tLO \nonumber
   \\&+ F_\tNLO+y_\tNLO \partial_y F_\tLO+\bar{\phi}_\tNLO \partial_\phi F_\tLO+\dots,
\end{align}
where $\bar{\phi}_\tNLO$ is given by $\bar{\phi}_\tNLO=\phi_\tNLO+y_\tNLO \partial_y \phi_\tLO$, and all terms are evaluated at $\phi_\tLO$, $y_\tLO$.

\begin{figure*}[t]
    \begin{center}
    \includegraphics[width=0.9\textwidth]{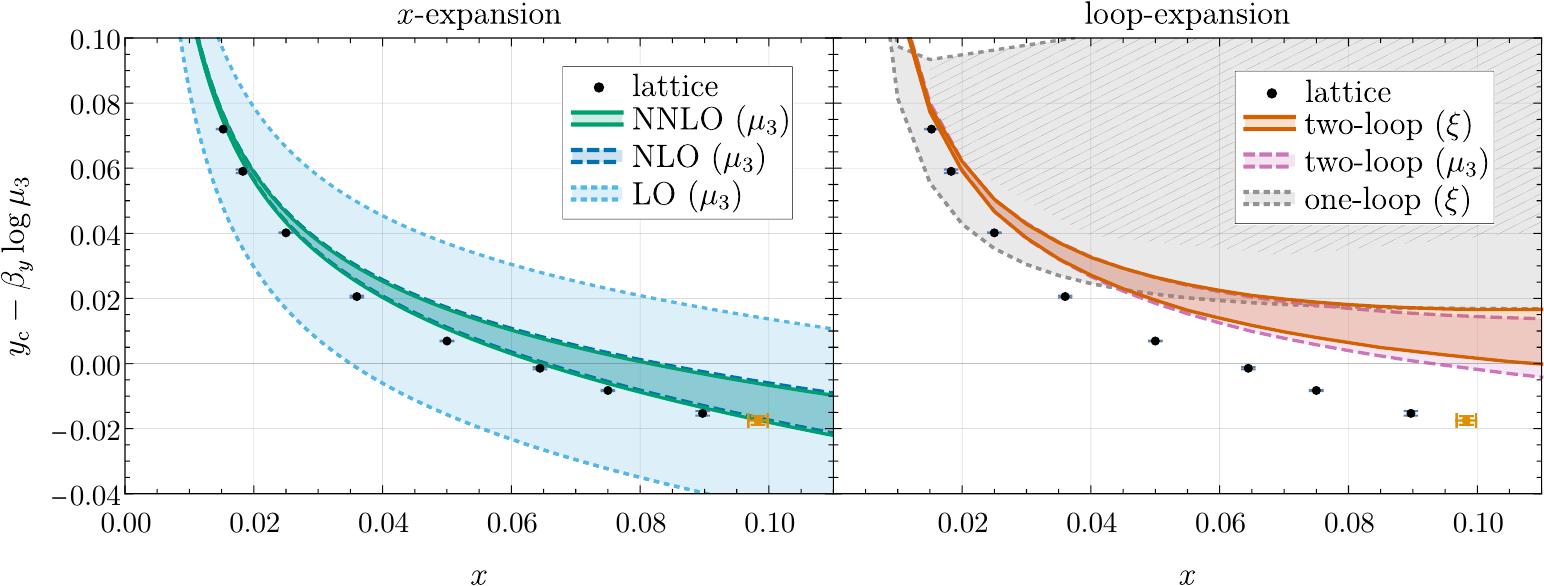}
    \end{center}
    \caption{The renormalization-scale invariant quantity $\yc-\beta_y\log\mu_3$, where $\yc$ is the critical mass, $\mu_3$ is the renormalisation scale, and $\beta_y$ is the beta function for $y$.
  Theoretical uncertainties for lattice data are shown as (barely visible) error bars, while those for the perturbative calculations are shown as coloured bands. The latter are estimated either by varying the renormalization scale over $\mu_3/(2 m_A) \in [1/\sqrt{10},\sqrt{10}]$, or by varying the gauge parameter over $\xi\in[0,5]$. The hashed region shown on the right reflects the presence of a third unphysical phase in this approximation.
    The orange data point shows the location of the endpoint of the line of first-order phase transitions, as determined on the lattice.}
    \label{fig:yc}
\end{figure*}

Note the simplicity of using strict-perturbation theory as compared to a numerical approach: there is no need to numerically minimize (the real parts of) complicated potentials since everything is expressed in terms of the LO vev and critical mass.

Let us now turn to the latent heat, which determines the strength of a first-order phase transition. It is given by the temperature derivative of $\Delta V$. This can be written in terms of derivatives with respect to the 3d effective parameters $x$, $y$ and $g_3^3$, using the chain rule of differentiation \cite{Farakos:1994xh}. Dependence on the derivative with respect to $g_3^3$ cancels at the critical temperature \cite{Kajantie:1995kf}. This leads us to calculate the following scalar condensates,
\begin{align} \label{eq:condensates}
    \Delta\braket{\hsq} &\define \frac{\partial}{\partial y} \Delta V,&
    \Delta\braket{(\hsq)^2} &\define \frac{\partial}{\partial x} \Delta V,
\end{align}
which determine the contribution to the latent heat from the infrared EFT scale.

\sectioninline{Results}%
Results at LO follow from the LO effective potential given in equation \eqref{eq:LeadingOrderPotential}.
The minima can be found analytically, which yields simple expressions for thermodynamic observables.

NLO corrections arise from two-loop diagrams with loop momenta of order the vector boson mass. These diagrams
appear through integrating out the heavy vectors for the EFT of the light scalars.
Contributions to these diagrams from smaller loop momenta are subdominant, so scalar masses can be set to zero within loop integrals. The total NLO contribution to the effective potential is
\begin{align}
    V_\tNLO=\frac{\phi ^2}{ (4\pi) ^2}  \left(-\frac{51}{32} \log \frac{|\phi|}{\mu_3} -\frac{63}{32} \log \frac{3}{2} + \frac{33}{64}\right),
\end{align}
where $\mu_3$ is the 3d renormalization scale \cite{Farakos:1994kx}.

The NNLO potential consists of one-loop diagrams within the light scalar EFT. 
These diagrams give
\begin{align}
    V_\tNNLO=-\frac{1}{12 \pi }\left[\left(m_H^2(\phi)\right)^{3/2} + 3 \left(m_G^2(\phi)\right)^{3/2}\right].
    \label{eq:VNNLO}
\end{align}
The resummation of vector petals follows from equation \eqref{eq:resummed_masses}.


With the effective potential in hand, we can calculate desired observables. Following the previous section we find
\begin{align}
    \yc &= \frac{
        1 - \frac{51}{2} x \log \tilde{\mu_3} - 2 \sqrt{2} x^{3/2}
    }{2 (8 \pi)^2 x},
    \label{eq:mcRes}\\
    \Delta\braket{\hsq}_\text{c} &= \frac{
        1 + \frac{51}{2} x + 13 \sqrt{2} x^{3/2}
    }{2 (8 \pi x) ^2},
    \label{eq:phisqRes}\\
    \Delta\braket{\left(\hsq\right)^2}_\text{c} &= \frac{
        1 + 51 x + 14 \sqrt{2} x^{3/2}
    }{4 (8\pi x)^4}.
    \label{eq:phisqsqRes}  
\end{align}
where $\tilde{\mu}_3\define e^{\tfrac{11}{34}-\tfrac{42}{34}\log\tfrac{3}{2}} (8\pi x \mu_3) \approx 0.84 (8 \pi x \mu_3)$. These expressions are accurate up to $O(x^2)$ in the numerators.

Some information on the convergence of the perturbative expansion can be gleaned by comparing the magnitudes of successive terms. One finds that NLO terms in equations \eqref{eq:mcRes}-\eqref{eq:phisqsqRes} can dominate over LO for $x\gtrsim 0.03$, choosing a renormalization scale $\mu_3\sim 1/(8\pi x)$. However, in each case the NNLO term remains smaller until significantly larger $x$. This suggests that the NLO term is anomalously large, and the perturbative expansion may be reliable at somewhat larger values of $x$, as indeed we find in our comparison to lattice data.

The explicit logarithm of the renormalization scale in equation \eqref{eq:mcRes} is crucial to cancel the implicit running of $y$ at that order; for the beta functions see Refs.~\cite{Farakos:1994kx, Gould:2022ran}. By contrast, the absence of logarithms in the scalar condensates reflects the renormalization group invariance of these physical quantities.

The $x$-expansion is gauge-invariant and renormalization scale invariant order-by-order. However, residual renormalization scale dependence can be useful to estimate theoretical uncertainties. For $\yc$, this can be carried out by including the exact renormalization group running, rather than truncating at the order of $x$ calculated. Doing so includes running from terms suppressed by higher powers of $x$ than we have calculated, which therefore do not cancel. For the scalar condensates this is not possible, as the quantities are simply renormalization group invariant, both exactly and order-by-order, so there is no way to introduce renormalization group running. This is nevertheless a desirable feature of the $x$-expansion, as it gives unambiguous predictions at each order.

\begin{figure}[t]
    \begin{center}
    \includegraphics[width=0.48\textwidth]{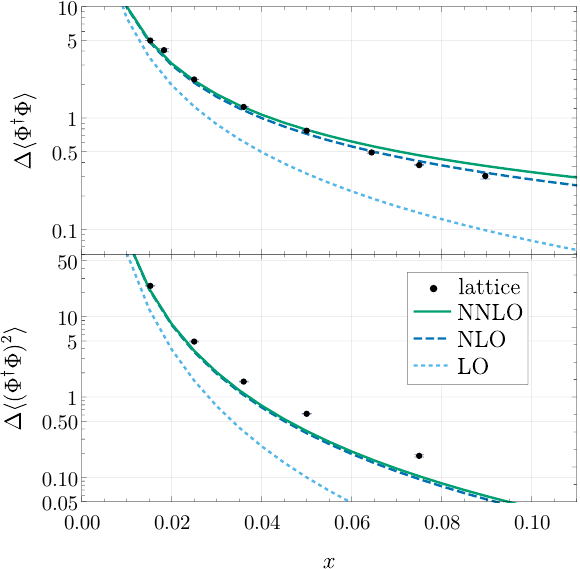}
    \end{center}
    \caption{The jumps in the scalar condensates as functions of $x$, computed in the $x$-expansion and on the lattice. Both condensates are manifestly gauge and renormalization-scale invariant.}
    \label{fig:condensates}
\end{figure}

Figures \ref{fig:yc} and \ref{fig:condensates} compare perturbative and lattice results, the latter taken from Refs.~\cite{Kajantie:1995kf, Gurtler:1997hr, Rummukainen:1998as, Laine:1998jb, Gould:2022ran}. The $x$-expansion for $\yc$ agrees well with the lattice over the entire range of $x$ for which there is a first-order phase transition, and the renormalization scale dependence gives a good estimate of the disagreement. There is a significant improvement from LO to NLO, and while the NNLO correction is small, it does shift towards the lattice data. Figure \ref{fig:condensates} shows similarly good agreement for the scalar condensates, though in this case the renormalization scale invariance of the $x$-expansion means there are no uncertainty bands. For the quadratic condensate, the NNLO correction improves agreement with the lattice only for $x \lesssim 0.05$, suggesting that above this higher-order terms become important. The quartic condensate shows the largest discrepancies between the $x$-expansion and the lattice data, which is consistent with the larger expansion coefficients for this quantity.

For comparison, in figure \ref{fig:yc} we also show predictions based on a different perturbative approach: numerically minimising the real part of the one-loop and two-loop effective potentials.
This approach is commonly taken in the literature but, for radiatively-induced transitions, it is not an expansion in any small parameter, because loop-level contributions are of the same size as tree-level terms. Further, directly minimising the real part of the effective potential cancels neither gauge nor renormalization-scale dependence at each order. The large uncertainty at one-loop order in this approach is due to the presence of a third unphysical phase for gauge parameters $\xi\gtrsim 3$. Predictions for all the condensates in this approach gain an imaginary part for $y<0$.

\sectioninline{Conclusions}%
We find that strong first-order phase transitions can be described rather well by perturbation theory.
We have developed an EFT approach to transitions with radiatively-generated barriers, and performed calculations up to NNLO.
The main results of this paper are presented in equations~\eqref{eq:mcRes}--\eqref{eq:phisqsqRes} and in Figs.~\ref{fig:yc} and \ref{fig:condensates}.
For the SU(2) Higgs theory, we find good numerical agreement between our perturbative expansion and existing lattice data.

The $x$-expansion is theoretically well-behaved in the sense that it does not suffer from IR-divergences or gauge-dependence, and residual renormalization scale-dependence is either absent (for the condensates), or is as sizable as expected (for the critical mass). In addition, contrary to previous calculations \cite{Arnold:1994bp, Arnold:1992rz}, our results indicate that the perturbative expansion converges. Though the results are consistent with lattice calculations, and better behaved than those of other methods, the expansion can still be improved. As we have seen, going to NLO is very important to get a reasonably accurate prediction. The next improvement to NNLO is numerically smaller, but improves agreement with the lattice results for $x\lesssim 0.05$.

Our results can be contrasted with QCD, where there is a sizeable mismatch between perturbative and lattice calculations of the free energy \cite{Zhai:1995ac, Braaten:1995jr}.
This can be partially ascribed to the Linde problem \cite{Linde:1980ts}, but also to the relatively large magnitude of the QCD gauge coupling, and its slow (logarithmic) approach to asymptotic freedom.
Though the Linde problem is also present in the model we have studied, it appears first at N$^5$LO in the $x$-expansion, suppressed by $O(x^3)$. Thus, the $x$-expansion performs well, at least for small enough $x$.

Future studies can utilize the approach taken here for other models. The remaining higher-order terms in the expansion should also be calculated: N$^3$LO and N$^4$LO, respectively $O(x^2)$ and $O(x^{5/2})$. Doing so requires calculation of three-loop diagrams. These two terms are the highest order terms calculable in perturbation theory and therefore give the final word on how well the expansion actually works.

Finally, given the importance of a strict expansion for equilibrium quantities, we expect similar behaviour for near-equilibrium ones, like the bubble-nucleation rate \cite{Ekstedt:2022ceo}. As such our results are an important stepping-stone for accurate predictions of gravitational waves.

\begin{acknowledgments}
We would like to thank Sinan G\"{u}yer and Kari Rummukainen for making available their lattice results prior to publication of Ref.~\cite{Gould:2022ran}.
We would also like to thank Tuomas Tenkanen for spotting a minus sign error in the first version.
The work of A.E. has been supported by the Deutsche Forschungsgemeinschaft under Germany's Excellence Strategy - EXC $2121$ Quantum Universe - $390833306$; and by the Swedish Research Council, project number VR:$2021$-$00363$.
O.G.~(ORCID 0000-0002-7815-3379) was supported by U.K.~Science and Technology Facilities Council (STFC) Consolidated grant ST/T000732/1.
\end{acknowledgments}

\bibliography{refs}

\end{document}